**Clock and Trigger Synchronization between Several Chassis of Digital Data Acquisition Modules**

W. Hennig, H. Tan, M. Walby, P. Grudberg, A. Fallu-Labruyere, W.K. Warburton,
XIA LLC, 31057 Genstar Road, Hayward, CA 94544, USA
C. Vaman, K. Starosta, D. Miller,
National Superconducting Cyclotron Laboratory, Michigan State University, East Lansing, MI 48824, USA

In applications with segmented high purity Ge detectors or other detector arrays with tens or hundreds of channels, where the high development cost and limited flexibility of application specific integrated circuits outweigh their benefits of low power and small size, the readout electronics typically consist of multi-channel data acquisition modules in a common chassis for power, clock and trigger distribution, and data readout. As arrays become larger and reach several hundred channels, the readout electronics have to be divided over several chassis, but still must maintain precise synchronization of clocks and trigger signals across all channels. This division becomes necessary not only because of limits given by the instrumentation standards on module size and chassis slot numbers, but also because data readout times increase when more modules share the same data bus and because power requirements approach the limits of readily available power supplies. In this paper, we present a method for distributing clocks and triggers between 4 PXI chassis containing DGF Pixie-16 modules with up to 226 acquisition channels per chassis in a data acquisition system intended to instrument the over 600 channels of the SeGA detector array at the National Superconducting Cyclotron Laboratory. Our solution is designed to achieve synchronous acquisition of detector waveforms from all channels with a jitter of less then 1 ns, and can be extended to a larger number of chassis if desired.

### 1. Introduction

In applications with segmented HPGe detectors or other detector arrays with tens or hundreds of channels (where the high development cost and limited flexibility of application specific integrated circuits outweigh their benefits of low power and small size), the readout electronics typically consist of multi-channel data acquisition modules in a common chassis for power, clock and trigger distribution, and data readout. As arrays become larger, the modules must be housed in several chassis, not only because of limits given by the instrumentation standards on module size and chassis slot numbers, but also because data readout times increase when more modules share the same data bus and because power requirements approach the limits of readily available power supplies. However, the system still must maintain precise synchronization of clocks and trigger signals across all channels. In this paper, we present a method to distribute clock and trigger signals in a new digital data acquisition system (DDAS) developed for the 18-detector SeGA array [1] at the National Superconducting Cyclotron Laboratory (NSCL).

The SeGA array's over 600 channels will be instrumented with 39 DGF-Pixie-16 modules, housed in four CompactPCI/PXI chassis. The Pixie-16 modules will be organized into a group of 3 trigger modules (the "director" and two "assistants", connecting to the detectors' central contacts) and 18 pairs of acquisition modules ("manager/worker" pairs connecting to the 32 segment channels from each detector). To reconstruct events throughout the detector system and perform gamma ray tracking analysis within a detector, the individual modules have to acquire detector waveforms and measure pulse heights simultaneously, i.e. any delays between modules must be small and stay constant throughout the data acquisition with minimal variations from event to event (jitter). In practice, simultaneous data acquisition requires the system to meet 3 conditions:

1.  Distribute a common reference clock to all modules in the system. This clock – in the present case a 50 MHz signal doubled on each module to 100 MHz – is the base for all digitization, signaling, and processing in the modules.
2.  Send global trigger signals from the director to all modules, using the coincidence pattern from the central contacts and several auxiliary detectors to limit data recording to only the events of interest. These trigger signals have to be distributed with tight timing constraints to all modules in the system, so that the assistants, managers, and workers acquire waveforms and pulse height measurements synchronously for the central contact and all 32 segments of a detector, in the present case with 1 ns jitter or less.
3.  Provide the director with a signal indicating that <u>all</u> modules are ready for data acquisition to avoid recording of partial events and/or corruption of data when individual modules are not ready, for example due to processing of previous events or to a backlog in data readout to the host computer. Obviously in practice the acquisition system will be designed to maximize the time it is ready, but nevertheless it must be able to determine when it is not ready and to cope with the consequences. The timing for this signal, which can be implemented as an AND of READY signals from individual modules, is relatively relaxed, on the order of microseconds.

The system of Pixie-16 modules, chassis, and P16Trigger boards described below provides the hardware to distribute clock and trigger signals according to the DDAS requirements. A 2-chassis test system has been set up and evaluated. The actual implementation of coincidence triggering and synchronous waveform acquisition in the firmware and software of the Pixie-16 is beyond the scope of this paper and will be reported elsewhere.

## 2. Experimental Setup

The DGF Pixie-16 is a 16-channel data acquisition module providing digital spectrometry and waveform acquisition. A block diagram is shown in Fig. 1. Similar in architecture and functionality to the DGF Pixie-4 [2], each channel of the Pixie-16 accepts signals directly from a detector preamplifier or photomultiplier tube. Signals are digitized in a 12-bit, 100 MHz ADC. Triggering, pile-up inspection and filtering of the data stream is performed in real time in field programmable gate arrays (FPGA); pulse heights and other event data are calculated on an event-by event basis in a digital signal processor (DSP). Results are stored in 512K of MCA memory and 128k of list mode FIFO memory. The Pixie-16 communicates with the host computer through a 32bit, 33MHz PCI interface which allows data transfer rates of up to 109MByte/s.

The Pixie-16 is operated in a 6U Compact PCI/PXI chassis with high current power supplies. The lower half of the backplane contains the PCI data bus interface compatible with the 3U CompactPCI/PXI standard, including a low-skew (<1ns) clock distribution path from a master clock in slot 2 to all slots in the chassis. The upper half of the backplane provides more than 160 lines for clock, trigger and data distribution between modules, as well as power connections to the module. A second card cage at the rear of the backplane allows rear I/O modules to also connect to the clock and trigger lines.

The P16Trigger module plugs into the chassis backplane from the rear, as indicated in Fig.2. When configured as a clock source module, it converts a clock signal from a Pixie-16 module (or from an on-board oscillator) into an LVDS signal and distributes it via CAT-5 cables to up to 8 P16Trigger modules configured as clock receivers. One of the clock receivers is the clock master module itself, ensuring minimal clock skew between chassis if CAT-5 cables of equal length are used. The incoming CAT-5 clock is then fed to the chassis clock distribution path via a Pixie-16 module in slot 2.

Using the same CAT-5 cables, a P16Trigger module can also send 3 (alternatively 6) trigger signals from the director to 8 (alternatively 4) receiver modules that transmit the trigger signals to their local backplanes and thus to the Pixie-16 manager/worker pairs in a chassis. Signaling can be reversed for any of the 3 (6) signals, which allows the P16Trigger modules to pick up a *chassis-wide* READY signal from each chassis's backplane, build an AND on the P16Trigger module in the director's chassis, and then send a *system-wide* READY signal to the director module itself.

**3. Results and Discussion**
Even using their standard firmware and software, Pixie-16 modules can synchronously acquire waveforms in all channels when operated in the same chassis. For the data shown in Fig. 3, identical signals from a fast pulser were connected to 33 channels on 3 Pixie-16 modules. One channel ("center") was configured to trigger acquisition of pulse waveforms in all other channels. The time of arrival *differences* $\Delta T_N$ between the center and each channel N were calculated for 700 sets of 33 waveforms ("events"). Histogramming $\Delta T_N$ of all events, we obtain an approximately Gaussian distribution for every channel. Fig. 3 shows the peak position and FWHM of each channel's histogram, demonstrating that the FWHM and thus the trigger jitter is less than 300ps for every channel. Though $\Delta T_N$ is ideally zero for all channels, in practice there may be a *fixed* time offset between channels due to cable delays etc. Thus the peak position of the Gaussian is shifted away from zero by a few nanoseconds. When the measurement is repeated with an additional cable delay in some channels, the peak position for these channel shifts by an additional offset corresponding to the cable length. Note that this offset is the same in each event and thus can be calibrated out at the beginning of a data acquisition.

Dedicated firmware and software for the SeGA application are currently still under development and so far the P16Trigger modules have only been tested with pulser signals. In the tests, the local oscillator of the P16Trigger was used as the clock source and clock signals probed on points C1-C6 (as indicated in Figure 2) with an oscilloscope. Figure 4 shows envelopes of clock signals acquired over several seconds (triggered on the upper curve). The width of the trace at the intersection is thus an estimate of *maximum* clock variations between two points over many millions of traces. Within the same module (a) and between two modules in the same chassis (b) this variation is less than 1 ns. Between the backplane's PXI clock input of two different chassis (equivalent to the P16Trigger output) the variation is about 1ns (c). Between two modules in different chassis (d) the variation is about 1.5 ns. Note that signals are distorted by reflections and pickup in the oscilloscope probes, and thus represent a worst case estimate.

In the SeGA application software, the director will issue several types of triggers to the manager/worker pairs. To test this function, a fast pulser was connected to point T1 on one P16Trigger module to simulate triggers issued by the director. Output signals were probed at T2 on the same P16trigger module and at T3 on a second P16Trigger module in a second chassis (see Fig. 2). As shown in Figure 4, the output signals both have a delay of ~18ns to the input signal, and a maximum skew of about 1ns between them.

In the SeGA application software, each manager/ worker will connect its module-wide READY signal to a common wire-OR line on the backplane, and one module will transmit the status of this line as the chassis-wide READY signal to the P16Trigger module. With the P16Trigger modules configured for reverse signaling, READY signals from all chassis are combined at the P16Trigger module in the director chassis, where the incoming reversed signals are combined in an AND gate. To test this function, signals from 2 fast pulsers were connected to points T2 and T3 on both P16Trigger modules and the output of the combined signals was probed at T1. As shown in Figure 5, the output signal is an AND of the two input signals. The signal delay from input to output is again about 18 ns for either input and/or edge.

## 4. Conclusion

In summary, a trigger module was built for distributing clocks and triggers between several chassis of Pixie-16 modules in the SeGA DDAS. While dedicated SeGA firmware/software is under development, the P16Trigger module was tested with local clocks and pulser signals. The P16Trigger transmits 50MHz clock signals with high quality and achieves a channel to channel trigger skew of less than 1ns. Clock variations within a chassis and between the outputs of the P16Trigger module are 1ns or less. Variations between modules in different chassis (after the backplane PXI clock buffer) were about 1.5ns, as measured using an oscilloscope. Note that this measurement of the maximum variation of any two clock pulses within several seconds is the worst case estimate of clock jitter and is in addition increased by reflections and distortions in the oscilloscope probes. It is *not* the variation between two momentary clock pulses which determine the trigger time for waveform capture, which we expect to be at least a factor of two smaller. When firmware/software are ready, measurements with Pixie-16 modules will be performed to determine if this clock variation increases the jitter in synchronous waveform acquisition, and if necessary the chassis clock distribution will be improved.

## 5. Acknowledgments

This material is based upon work supported by the US National Science Foundation under Grants No. MRI-0420778 and Phy-0606007.

# 5. Figures

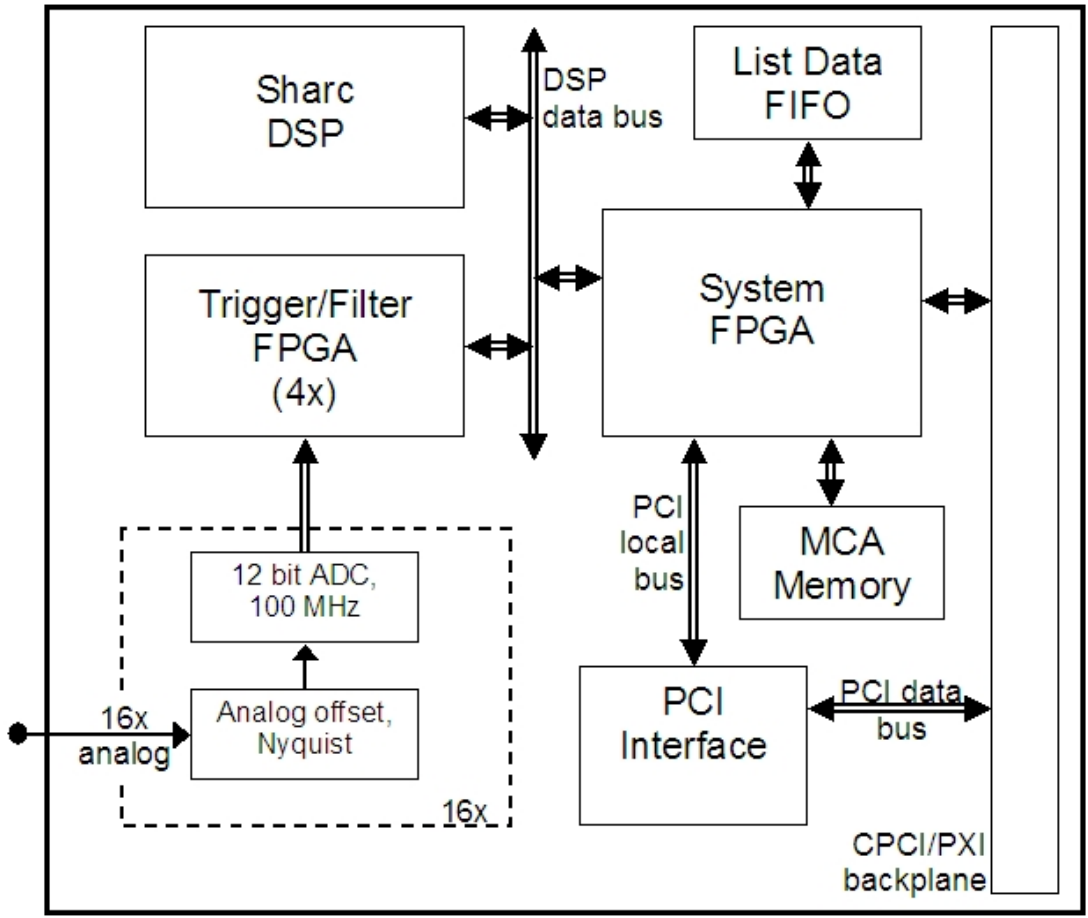

**Figure 1: Simplified block diagram of the DGF Pixie-16**

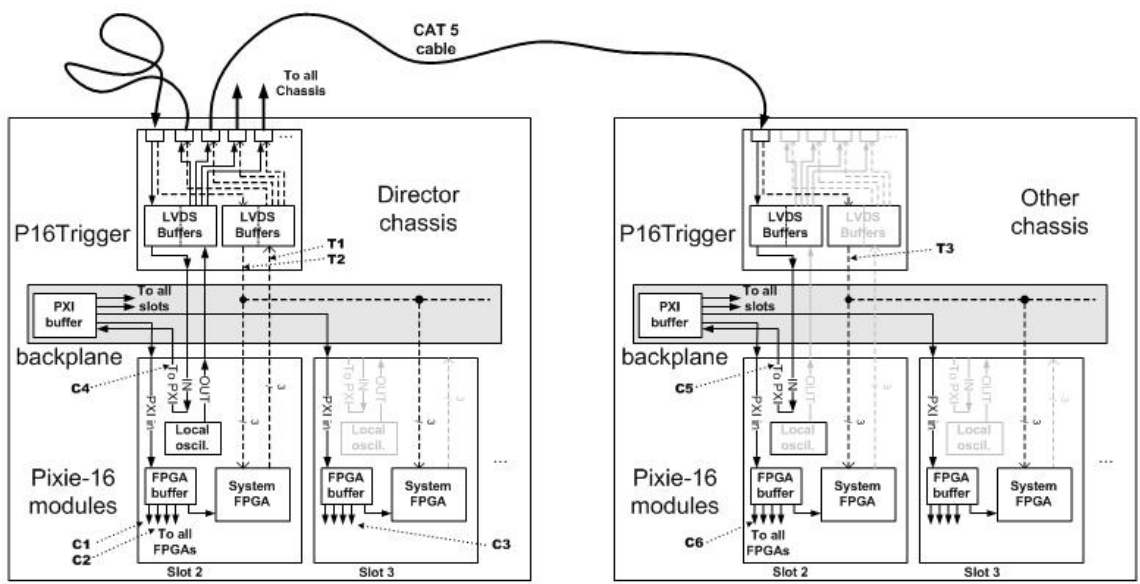

**Figure. 2: Distribution of clock (solid lines) and triggers (dashed) with P16Trigger modules.**

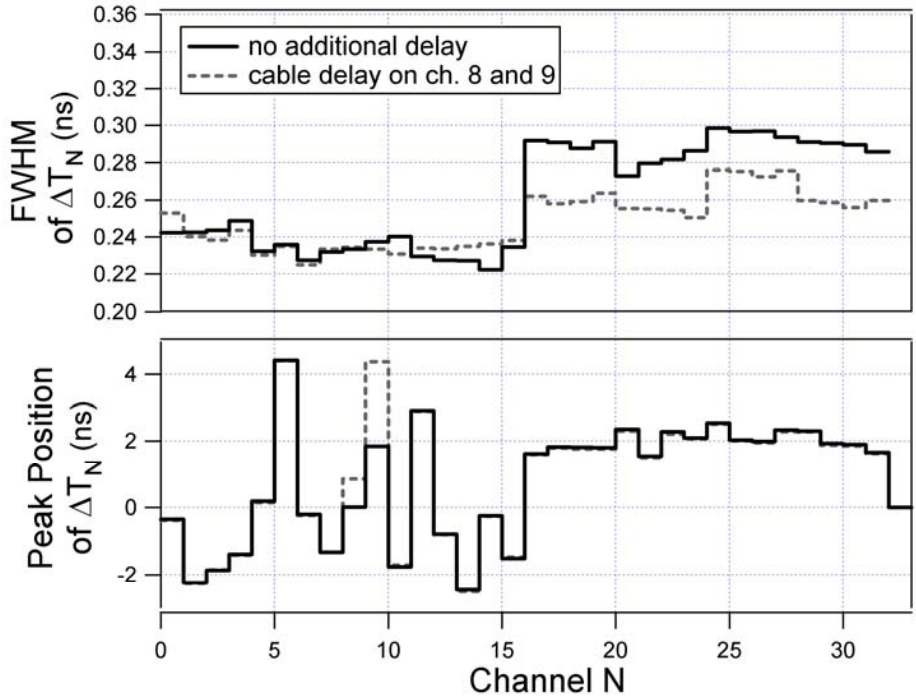

**Figure 3: Peak position and FWHM of time of arrival difference $\Delta T_N$ between "center" channel and 32 channels**

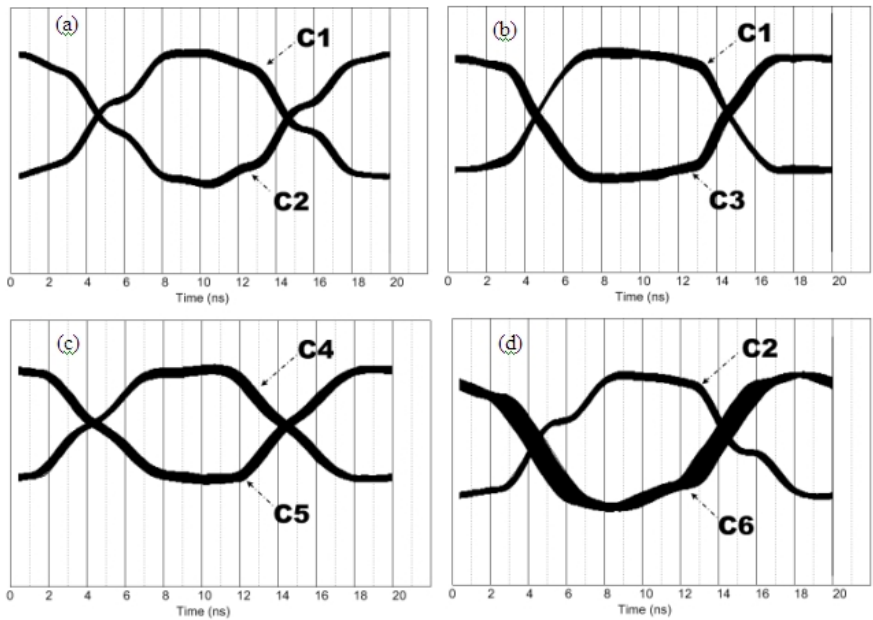

**Figure 3: Oscilloscope screen shots from clock distribution tests.**

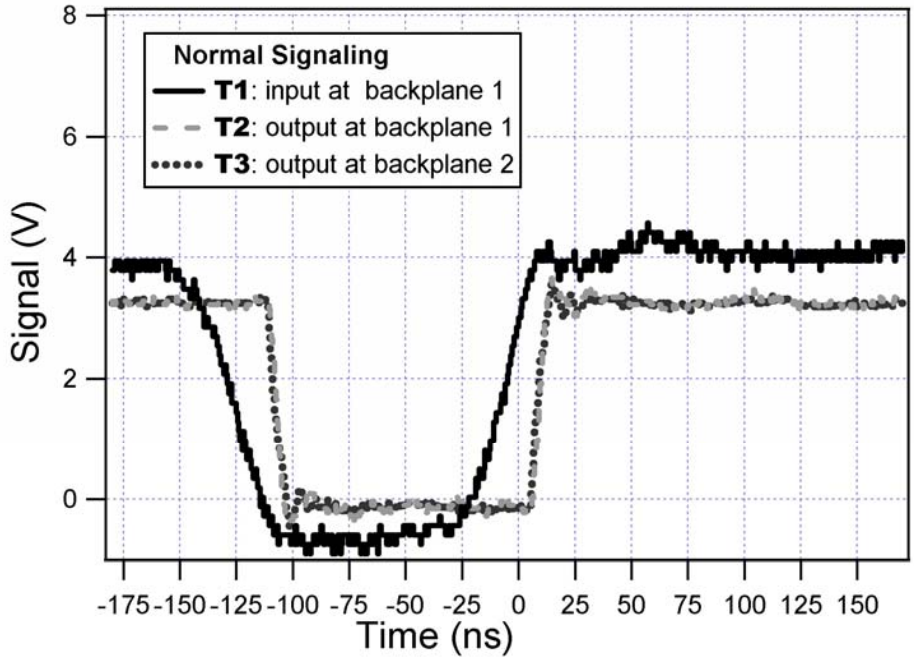

**Figure 4: Trigger signals issued from the director chassis to both chassis.**

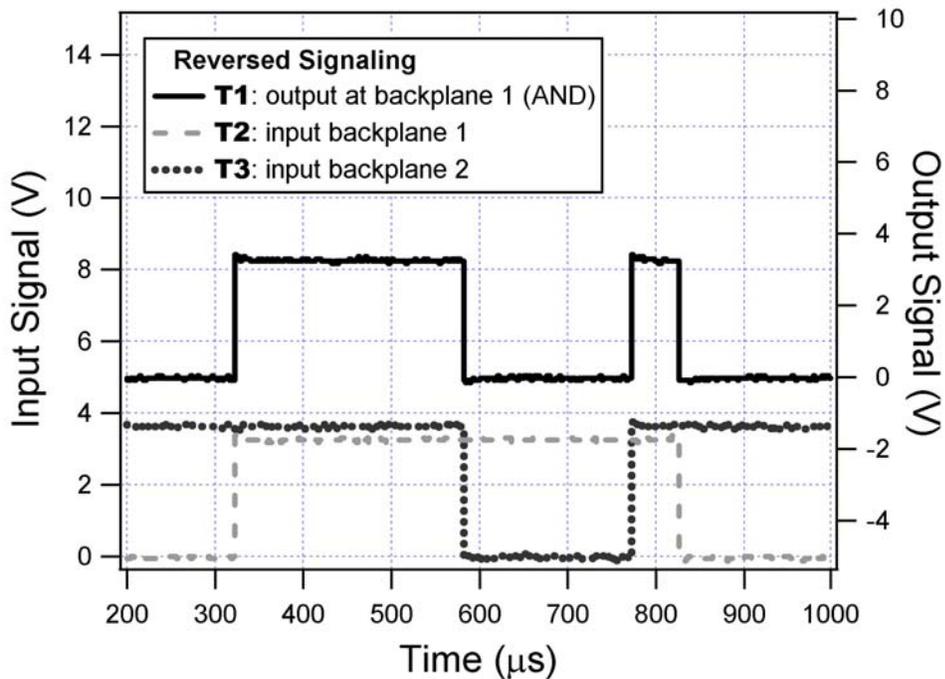

**Figure 5: READY signals from both chassis are combined to system-wide READY by forming an AND on the P16Trigger module in the director chassis.**